\documentclass[12pt]{iopart}
\usepackage{iopams}

\usepackage{graphicx,bm}
\usepackage{txfonts}
\usepackage{color}
\usepackage{hyperref}
\newcommand{\rp}[1]{(\ref{#1})}

\newcommand{\abs}[1]{\left|{#1}\right|}

\newcommand{\av}[1]{\left\langle #1 \right\rangle}

\newcommand{\bbr}[1]{\langle #1|}

\newcommand{\ke}[1]{|#1\rangle}

\newcommand{\wt}[0]{\widetilde}
\newcommand{\oo}[0]{^{\circ}}
\newcommand{\al}[1]{^{(#1)}}
\newcommand{\da}{^\dagger}

\newcommand{\ppt}[1]{\left( #1 \right)}
\newcommand{\pq}[1]{\left[ #1 \right]}
\newcommand{\pg}[1]{\left\{ #1 \right\}}

\newcommand{\ee}{{\rm e}}

\usepackage{bbold}
\newcommand{\id}{\mathbb{1}}%{\openone}%{\mathbbm{1}}

\newcommand{\nn}{{\nonumber}}

\newcommand{\mmat}[2]{
                      \begin{array}{#1}
                       #2
                       \end{array}  }

\newcommand{\ovl}{\overline}

\newcommand{\va}{{\bf a}}

\newcommand{\vf}{{\bf f}}

\newcommand{\vx}{{\bf x}}

\newcommand{\AAA}{{\cal A}}
\newcommand{\BBB}{{\cal B}}
\newcommand{\CC}{{\cal C}}

\newcommand{\GG}{{\cal G}}

\newcommand{\KK}{{\cal K}}

\newcommand{\LL}{{\cal L}}
\newcommand{\MM}{{\cal M}}
\newcommand{\NN}{{\cal N}}
\newcommand{\OO}{{\cal O}}

\newcommand{\RR}{{\cal R}}
\newcommand{\TT}{{\cal T}}

\newcommand{\VV}{{\cal V}}
\newcommand{\WW}{{\cal W}}
\newcommand{\XX}{{\cal X}}
\newcommand{\YY}{{\cal Y}}
\newcommand{\ZZ}{{\cal Z}}

\definecolor{blue}{rgb}{0,0,0.8}

\definecolor{green}{rgb}{0,0.6,0}

\usepackage[normalem]{ulem}
\newcommand{\stkout}[1]{\ifmmode\textrm{\sout{\ensuremath{#1}}}\else\sout{#1}\fi}

\begin{document}

\title{
Generation of stable Gaussian cluster states in optomechanical systems with multifrequency drives
}

\author{Nahid Yazdi$^{1,2}$, Stefano Zippilli$^{1,*}$, David Vitali$^{1,3,4}$}
\address{$^1$ School of Science and Technology, Physics Division, University of Camerino, I-62032 Camerino (MC), Italy\\
$^2$ Department of Physics, Isfahan University of Technology, Isfahan 84156-83111, Iran\\
$^3$ INFN, Sezione di Perugia, I-06123 Perugia, Italy\\
$^4$ CNR-INO, I-50125 Firenze, Italy\\
$^*$ Corresponding author}
\ead{stefano.zippilli@unicam.it}

\vspace{10pt}
\begin{indented}
\item[]\today
\end{indented}

\begin{abstract}
We show how to dissipatively stabilize the quantum state of $N$ mechanical resonators in an optomechanical system, where the resonators interact by radiation pressure with $N$ optical modes, which are driven by properly selected multifrequency drives. We analyze the performance of this approach for the stationary preparation of Gaussian cluster states.
\end{abstract}

\section{Introduction}\label{Intro}

Over the last decade, an increasing number of experiments have explored the potentiality of multimode optomechanics, where various vibrational modes of mechanical resonators interact, in a controlled way, with one or more electromagnetic field modes~\cite{
massel2012,
fan2015,
barzanjeh2017,
bernier2017a,
peterson2017a,
kralj2017,
nielsen2017,
nair2017,
weaver2017,
gil-santos2017,
ruesink2018,
piergentili2018,
moaddelhaghighi2018,
gartner2018a,
riedinger2018,
ockeloen-korppi2018,
colombano2019,
mathew2020,
sheng2020,
piergentili2021,
fiaschi2021,
kotler2021,
mercierdelepinay2021,
delpino2022,
jong2022,
kharel2022,
youssefi2022a,
ren2022a,
mercade2023%
}.
These investigations have been spurred by a number of theoretical proposals predicting many novel, interesting collective dynamics~\cite{
bhattacharya2008,chang2011,tomadin2012,xuereb2012%
}, 
including
non-reciprocal transport~\cite{metelmann2015a,
eshaqi-sani2022},
enhanced quantum sensing~\cite{brady2023},
quantum many-body effects~\cite{ludwig2013,ludwig2013,
peano2015,schmidt2015a,xuereb2014a,qi2019,lai2022a},
non-linear processes~\cite{heinrich2011a,holmes2012,
li2020,pelka2020,pelka2022,zhang2022},
and
mechanical entanglement~\cite{
tan2013,tan2013a,woolley2014,abdi2015,
li2015,
zippilli2015,
houhou2015a,li2017,asjad2016a,lai2022a}.
Multimode optomechanics has also been suggested as a platform for quantum information processing and computation~\cite{
houhou2015a, moore2016, moore2017a,houhou2022%
}.
Therefore, it is important to assess the requirements for achieving full control of the mechanical dynamics in such systems. 

Here, we employ reservoir engineering techniques, based on the use of multifrequency drives, to control the stationary quantum properties of many mechanical resonators. 
This approach builds upon several previous results regarding the entanglement of mechanical resonators%
~\cite{
tan2013,
tan2013a,
woolley2014,
abdi2015,
li2015,
houhou2015a,
zippilli2015,
li2017,
ockeloen-korppi2018,
mercierdelepinay2021,
kotler2021%
}. In the case of two mechanical modes, these approaches work by driving the system with laser fields resonant with both the red and blue mechanical sidebands, such that specific collective Bogoliubov modes can be cooled, resulting in two-mode mechanical squeezing, that is, entanglement.

In the present work, we study a multimode optomechanical system and 
show that in order to achieve full steady-state control of $N$ mechanical resonators, one needs $N$ optical modes, each driven on both the red and blue sidebands corresponding to all the mechanical resonators (see Fig.~\ref{fig1}). This requires the use of drivings with $2\,N^2$ frequency components. 
Thereby, the mechanical resonators can be laser-cooled to any Gaussian state, depending on the corresponding driving amplitudes and phases.
In particular, we analyze in detail the conditions for the stationary generation of Gaussian cluster states~\cite{zhang2006,zippilli2020, zippilli2021}, which constitute the fundamental resource of measurement-based quantum computation in the continuous variable setting~\cite{menicucci2006,gu2009}. 
Differently from the more common approach of gate-based quantum computation, where a sequence of unitary quantum operations is applied on a quantum processor, in measurement-based quantum computation, a quantum algorithm is realized by performing a sequence of measurements over certain highly entangled quantum states. These are the so-called cluster states, characterized by graphs that identify the structure of the entanglement between the components of the quantum processor. Therefore, in this approach, a fundamental part of the computation is the preparation of the cluster state. In the continuous variable setting, very large Gaussian cluster states of optical fields have been demonstrated~\cite{pfister2019}. Here, we suggest a path to achieving similar results with mechanical systems, thereby providing a step forward towards the realization of quantum computation over mechanical degrees of freedom~\cite{moore2017a,houhou2022}.

Our approach is somewhat similar to that of Ref.~\cite{houhou2015a} where, however, a single cavity mode is considered, so that the protocol requires the sequential application of specific driving pulses. In our proposal, on the contrary, the state preparation is truly stationary. In this respect, our result is analogous to the scheme proposed in Ref.~\cite{yazdi2023}, which shows how to stabilize the state of many microwave modes. 
 
The article is organized as follows. 
In section \ref{Model_BasicDyn}, we introduce the system, discuss the approximations at the basis of the analytical model, and analyze the conditions under which it is possible to have full control of the mechanical steady state. Then, in section~\ref{Cluster}, we specialize these results to the dissipative preparation of Gaussian cluster states. The performance of the scheme is analyzed in the next section~\ref{Results}. Finally, in section~\ref{Conclusions}, we draw our
conclusions. In the appendices, we report additional details about the derivation of the analytical model (\ref{Model}), and of the corresponding steady state (\ref{stst}).

\section{Model and basic dynamics}\label{Model_BasicDyn}

We consider an optomechanical system that involves the interaction of $N$ resonant modes of a mechanical object
(as, for example, in Refs.~\cite{
barzanjeh2017,
bernier2017a,
peterson2017a,
nielsen2017,
kralj2017,
mercade2023}) at frequencies $\Omega_j$ and $N$ resonant modes of an optical cavity at the frequencies $\omega_k$, for $k, j \in \{1, \ldots, N\}$. 
In this system, each cavity mode interacts with all the mechanical modes with strengths $g_{kj}$.
The system dynamics are controlled by driving the cavity with laser fields constituted of $2N^2$ spectral components at the frequencies $\lambda_{km}$, where $k \in \{1, \ldots, N\}$ and $m \in \{1, \ldots, 2N\}$.
The frequencies associated with the same index $k$ are chosen to be close to resonance with the frequency $\omega_k$ of an optical mode, that is, $\abs{\lambda_{km} - \omega_k } \ll \omega_k$ for all $m$. As a result, the components of the drives with index $k$ effectively drive only the corresponding cavity mode $k$. The multifrequency driving strength is given by
\begin{eqnarray}
\epsilon_k(t) = \sum_{m=1}^{2\,N} \ovl\epsilon_{km} e^{-i\,\lambda_{km}\ t} 
\ ,
\end{eqnarray}
where $\ovl\epsilon_{km}$ is the complex amplitude of the spectral component at frequency $\lambda_{km}$.

\begin{figure*}[t!]
\centering
\includegraphics[width=0.8\textwidth]{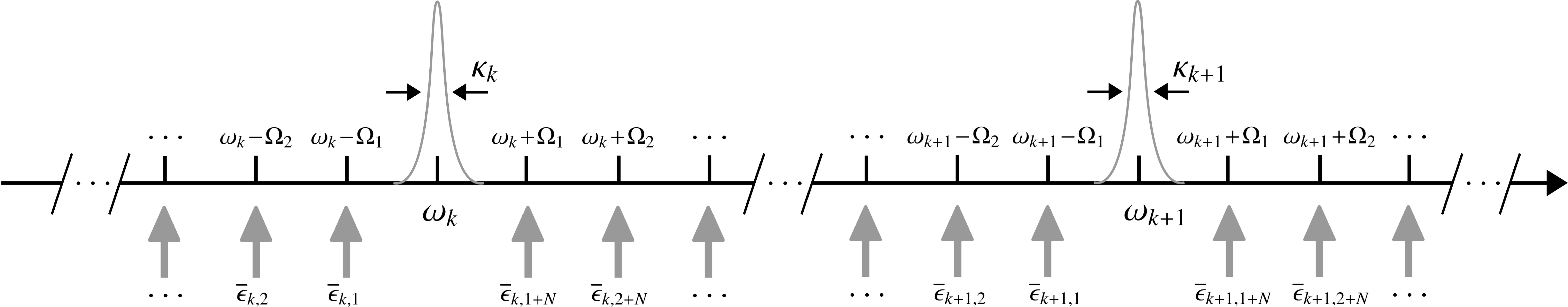}
\caption{Frequency scheme. $N$ optical modes at frequencies $\omega_k$ and linewidth $\kappa_k$, for $k\in\pg{1,\dots N}$, interact with $N$ mechanical modes at frequencies $\Omega_j$, for $j\in\pg{1,\dots N}$. The system is driven by multifrequency drives (upwards gray arrows) resonant with all the red and blue sidebands of all optical and mechanical modes, and with strengths $\ovl\epsilon_{k,j}$ and $\ovl\epsilon_{k,j+N}$.
}
\label{fig1}
\end{figure*}

As is customary with optomechanical systems, we consider the fluctuations of the system variables 
around the corresponding average values~\cite{aspelmeyer2014}. These fluctuations are
described by the bosonic annihilation operators $\wt a_k$ for the optical modes and $\wt b_j$ for the mechanical modes.
As detailed in \ref{Model}, the equations for the fluctuations can be linearized for large values of the average fields. In particular, simple equations can be obtained by assuming sufficiently small couplings $g_{kj}$ 
and expanding the average fields at the lowest relevant order in $g_{kj}$. 
Thereby, including dissipation and noise of the optical and mechanical modes at rates $\kappa_k$ and $\gamma_j$, respectively, the quantum Langevin equations for the slowly varying operators defined by $\wt a_k=e^{-i\,\wt\omega_k\ t}\,a_k$ [where $\wt\omega_k$ includes the frequency shift due to the optomechanical interaction, see Eq.~\rp{deltak}]  and $\wt b_j=e^{-i\,\Omega_j\ t}\,b_j$, take the form
(see \ref{Linearization})
\begin{eqnarray}\label{dotadotb}
\dot{a}_k &=& \ -\frac{\kappa_k}{2}\,a_k 
%- 2\,i\sum_{j=1}^{N}g_{kj}\ {\rm Re}\pq{\beta_j'(t)}a_k 
- i\sum_{j=1}^{N}\sum_{m=1}^{2\,N}\ 
g_{kj}\ \ovl\alpha_{km}
%\nn\\& &\ \times
\pq{
b_j\ e^{i\ppt{\wt\omega_k-\Omega_j-\lambda_{km}} t}+
b_j^\dagger\ e^{i\ppt{\wt\omega_k+\Omega_j-\lambda_{km}} t}
} 
+ \sqrt{\kappa_k}\ a_k^{in} 
\nn\\
\dot{b}_{j} &=&  -\frac{\gamma_j}{2}\,\wt b_j - i\sum_{k=1}^{N}\sum_{m=1}^{2\,N}\ 
g_{kj}
%\nn\\& &\ \times
\pq{
\ovl\alpha_{km}\ a_k^\dagger\ e^{i\ppt{\wt\omega_k+\Omega_j-\lambda_{km}} t} +
\ovl\alpha_{km}^*\ a_k\ e^{-i\ppt{\wt\omega_k-\Omega_j-\lambda_{km}} t}
} 
\nn\\& &
+ \sqrt{\gamma_j}\ b_j^{in} \ ,
\end{eqnarray}
where
\begin{eqnarray}\label{alphakm}
\ovl \alpha_{km}&=& \ \frac{\ovl\epsilon_{km}}{\lambda_{km}-\omega_k+i\frac{\kappa_k}{2}} \ ,
\end{eqnarray}
and where $a_k^{in}$ and $b_j^{in}$ are the input noise operators characterized by the correlation functions 
$\av{a_k^{in}(t)\ a_{k'}^{in}{}\da(t')}=\delta_{k,k'}\ \delta(t-t')$%
~\footnote{
Here and in the following the symbol $\delta_{k,k'}$ indicates a Kronecker delta while $\delta(t-t')$ a Dirac delta.
}, 
$\av{a_k^{in}{}\da(t)\ a_{k'}^{in}(t')}=\av{a_k^{in}{}\da(t)\ a_{k'}^{in}{}\da(t')}=\av{a_k^{in}(t)\ a_{k'}^{in}(t')}=0$, 
$\av{b_j^{in}(t)\ b_{j'}^{in}{}\da(t')}=\delta_{j,j'}\ \delta(t-t')\ppt{n_j+1}$, $\av{b_j^{in}{}\da(t)\ b_{j'}^{in}(t')}=\delta_{j,j'}\ \delta(t-t')\ n_j$ and $\av{b_j^{in}{}\da(t)\ b_{j'}^{in}{}\da(t')}=\av{b_j^{in}(t)\ b_{j'}^{in}(t')}=0$, with
$n_j$ the thermal phonon number at temperature $T$ defined by
\begin{eqnarray}
n_j=\frac{1}{e^{\hbar\,\Omega_j/k_B\,T}-1}\ . 
\end{eqnarray}

The driving frequencies can be selected in order to make resonant only those specific terms that allow for full control of the mechanical dynamics, while the effects of the residual time-dependent non-resonant terms are negligible  whenever their effective coupling strengths are much smaller than the corresponding frequencies. To be specific, we select, for $m\leq N$,
\begin{eqnarray}\label{lambda}
\lambda_{k,m}&=& \ \wt\omega_k-\Omega_m
\nn\\
\lambda_{k,m+N}&=& \ 
\wt\omega_k+\Omega_m
\end{eqnarray}
such that each cavity mode is driven both on the red and blue mechanical sidebands corresponding to each mechanical resonator (see Fig.~\ref{fig1}). We also assume that the system is in the resolved sideband regime~\cite{aspelmeyer2014} and that all the mechanical frequencies are sufficiently different from each other so that (see also \ref{condRWA})
\begin{eqnarray}\label{RWA-}
\kappa_j,\abs{g_{k,j}\ \ovl\alpha_{k,m}}, 
\abs{g_{k,j}\ \ovl\alpha_{k,m+N}}&\ll&\ \abs{\Omega_j-\Omega_m} 
\ \ \ \ \ \ \ \   {\rm for}\ j\neq m \ .
\end{eqnarray}
This relation justifies a rotating-wave approximation under which all the non-resonant terms are neglected.
Consequently, the quantum Langevin equations can be further approximated and rewritten as follows
\begin{eqnarray}\label{Eqab}
\dot{a}_k &=& \ -\frac{\kappa_k}{2}\,a_k 
- i\ \wt g_k\sum_{j=1}^{N}\
\pq{\XX_{k,j}\ b_j+
\YY_{k,j}\ b_j^\dagger\ 
} 
+ \sqrt{\kappa_k}\ a_k^{in} 
\nn\\
\dot{b}_{j} &=& \ -\frac{\gamma_j}{2}\,b_j - i\sum_{k=1}^{N}\ \wt g_k
\pq{
\YY_{k,j}\ a_k^\dagger +
\XX_{k,j}^*\ a_k
} 
+ \sqrt{\gamma_j}\ b_j^{in} \ ,
\end{eqnarray}
where we have introduced the matrices of interaction coefficients
\begin{eqnarray}\label{XXYYalpha}
\XX_{k,j}&=& \frac{g_{k,j}\ \ovl\alpha_{k,j}}{\wt g_k}
\nn\\
\YY_{k,j}&=& \frac{ g_{k,j}\ \ovl\alpha_{k,j+N} }{\wt g_k}\ ,
\end{eqnarray}
and the collective coupling strengths 
\begin{eqnarray}\label{tildeg2-}
\wt g_k=\sqrt{
\sum_{j=1}^N\ g_{kj}^2\,\pq{
\abs{\ovl\alpha_{kj}}^2
-\abs{\ovl\alpha_{k,j+N}}^2
}}
\ .
\end{eqnarray}
When the strengths and phases of the driving fields, $\ovl\epsilon_{km}$, are properly selected such that [see Eq.~\rp{alphakm} and \rp{XXYYalpha}]
\begin{eqnarray}\label{XXYY}
\XX\ \XX\da-\YY\ \YY\da&=& \ \id
\nn\\
\XX\ \YY^T-\YY\ \XX^T&=& \ 0\ ,
\end{eqnarray}
then these matrices define a Bogoliubov transformation, and the corresponding collective mechanical Bogoliubov modes
\begin{eqnarray}\label{cj-}
c_j=\sum_{j'=1}^N\ppt{\XX_{jj'}\ b_{j'}+
\YY_{j,j'}\ b_{j'}^\dagger}\ ,
\end{eqnarray}
fulfill the equations 
\begin{eqnarray}\label{QLEac-}
\dot{a}_k &=& \ -\frac{\kappa_k}{2}\,a_k 
- i\ \wt g_k\ c_k+
\sqrt{\kappa_k}\ a_k^{in} 
\nn\\
\dot c_k&=& \ -\sum_{k'=1}^N\ppt{ \WW_{k,k'}\ c_{k'} +\TT_{k,k'}\ c_{k'}\da} -i\ \wt g_k\ a_k+f_k
%\nn
\end{eqnarray}
where
\begin{eqnarray}\label{WT}
\WW_{k,k'}&=& \ \sum_{j=1}^N\frac{\gamma_j}{2}\ppt{\XX_{k,j}\ \XX_{k',j}^*-\YY_{k,j}\ \YY_{k',j}^*}
\nn\\
\TT_{k,k'}&=& \ \sum_{j=1}^N\frac{\gamma_j}{2}\ppt{\XX_{k,j}\ \YY_{k',j}-\YY_{k,j}\ \XX_{k',j}}\ ,
\end{eqnarray} 
and the collective noise operators are
\begin{eqnarray}\label{fk-}
f_k&=& \ \sum_{j=1}^N\ \sqrt{\gamma_j}\ppt{\XX_{k,j}\ b_j^{in}+\YY_{k,j}\ b_j^{in}{}\da}\ .
\end{eqnarray} 
In particular, we note that if all the mechanical dissipation rates are equal, i.e. $\gamma_j=\gamma$ for all $j$, then, according to Eq.~\rp{XXYY}, $\WW=
\frac{\gamma}{2}\
\id
$ and $\TT=0$.

Eq.~\rp{QLEac-} describes the laser cooling of all the mechanical Bogoliubov modes $c_j$. If the values of $\gamma_j$ (i.e., the couplings to the thermal reservoir) are negligible, these modes are cooled to the vacuum of modes $c_j$, characterized by the correlation functions 
\begin{eqnarray}\label{avcc}
\av{c_j\ c_{j'}\da}&=& \ \delta_{j,j'}
\nn\\
\av{c_j\da\ c_{j'}}&=& \ \av{c_j\ c_{j'}}=\av{c_j\da\ c_{j'}\da}= 0\ .
\end{eqnarray}
The corresponding result for the original modes (when $\gamma_j\to 0$) reads
\begin{eqnarray}\label{bb}
\av{b_j\ b_{j'}\da}&=& \ 
\sum_{k=1}^N\ \XX_{k,j}^*\ \XX_{k,j'}
\nn\\
\av{b_j\da\ b_{j'}}&=& \ \sum_{k=1}^N\ \YY_{k,j}^*\ \YY_{k,j'}
\nn\\
\av{b_j\ b_{j'}}&=& \ -\sum_{k=1}^N\ \XX_{k,j}^*\ \YY_{k,j'}
\nn\\
\av{b_j\da\ b_{j'}\da}&=& \ -\sum_{k=1}^N\ \YY_{k,j}^*\ \XX_{k,j'}\ .
\end{eqnarray}
These expressions are equal to the correlations evaluated over a pure Gaussian state $\ke{\psi}$ (such as $\bbr{\psi}b_j\,b_{j'}\da\ke{\psi}$) defined in terms of a Gaussian unitary operator $U$ applied to the vacuum of the $b$ modes
\begin{eqnarray}\label{psi-}
\ke{\psi}=U\,\ke{0}\ ,
\end{eqnarray}
when $U$ is characterized by the relation [see Eq.~\rp{cj-}]
\begin{eqnarray}\label{UbUc}
U\ b_j\ U\da=c_j\ ,
\end{eqnarray}
such that the inverse transformation is given by
\begin{eqnarray}\label{UdabU-}
U\da\,b_j\,U=\sum_{j'=1}^N
\ppt{\XX_{j',j}^*\ b_{j'}-\YY_{j',j}\ b_{j'}\da}\ .
\end{eqnarray}
This means that, by properly tuning the interaction strengths, which constitute the matrices $\XX$ and $\YY$, it is possible to drive the system to pure stationary Gaussian states of the form of Eq.~\rp{psi-}.
In particular, since we have no limitations over the matrices $\XX$ and $\YY$, which determine the Bogoliubov transformation, this scheme allows for the dissipative preparation of any pure zero-average Gaussian state of the mechanical modes. 
In other terms, this scheme realizes a sort of laser cooling to any collective mechanical zero-average Gaussian state.  

In the next sections, we apply this result to the dissipative preparation of  Gaussian cluster states.

\section{Stabilization of cluster states}\label{Cluster}

Let us now study the performance of our scheme for the preparation of a Gaussian cluster state with an adjacency matrix $\AAA$ (symmetric and with $\AAA_{j,j}=0$). It is defined as the state generated by applying the operator~\cite{zhang2006,zippilli2020, zippilli2021,menicucci2006,gu2009} 
\begin{eqnarray}\label{C}
C=\prod_{j,k=1}^N\ e^{\frac{i}{4}\ \AAA_{j,k}\ x_j\,x_k},
\end{eqnarray}
 with $x_j=b_j+b_j\da$, 
to a set of bosonic modes each in the zero eigenstate, $\ke{0}_p$, of the momentum operators \mbox{$p_j=-i\,b_j+i\,b_j\da$}, i.e. $\ke{\psi_{cluster}}=C\ \ke{0}_p$.  These are infinitely squeezed states and,
in practice, one can prepare approximate cluster states with finite squeezing. Here we study the preparation of states of the form 
\begin{eqnarray}\label{cluster-}
\ke{\psi_r}=C\ S_r\ \ke{0}
\end{eqnarray}
where $S_r$ is the squeezing operator $S_r=\ee^{\frac{r}{2}\sum_j\ppt{{b_j\da}^2-b_j^2}}$, and $\ke{0}$ is the vacuum. This state approaches an ideal cluster state in the limit of infinite squeezing, i.e. $\ke{\psi_{cluster}}=\lim_{r\to\infty}\ \ke{\psi_r}$.

The model of the previous section stabilizes a cluster state~\rp{cluster-} when 
the unitary matrix that generates the transformation in Eq.~\rp{UdabU-} is 
\begin{eqnarray}\label{CSU}
U=C\ S_r\ .
\end{eqnarray}
Using this relation, together with Eqs.~\rp{alphakm}, \rp{XXYYalpha}, \rp{cj-} and \rp{UbUc}, one can determine  the strength and phases of the driving fields for the stabilization of a cluster state. In detail, first, we compute how the unitary $C\ S_r$ transforms the operator $b_j$. And we  find
\begin{eqnarray}
S_r\da\ C\da\ b_j\ C\ S_r& &=c_r\ b_j+s_r\ b_j\da+\frac{i}{2}\ e^{r}\ \sum_{j'=1}^N\ \AAA_{j,j'}\ \ppt{b_{j'}+b_{j'}\da}
\end{eqnarray}
with $c_r=\cosh(r)$ and $s_r=\sinh(r)$ (where we used the relations 
$C\da\ b_j\ C=b_j+\frac{i}{2}\ \sum_{j'=1}^N\ \AAA_{j,j'}\ppt{b_{j'}+b_{j'}\da}$ and $S_r\da\ b_j\ S_r=c_r\ b_j+s_r\ b_j\da$
). Then, 
we compare this result to Eq.~\rp{UdabU-}, observing that the two are 
equal,
meaning the steady state of our system approximates a cluster state, when the following expressions are satisfied
\begin{eqnarray}\label{drivings-conditions-}
\XX&=& \ c_r\ \id-\frac{i}{2}\ e^r\ \AAA\ ,
\nn\\
\YY&=& \ -s_r\ \id-\frac{i}{2}\ e^r\ \AAA
\ .    
\end{eqnarray}
Finally, 
given any value of $r$ and any matrix $\AAA$ (which determine the expected cluster state), and any set of real positive values of $\wt g_k$ (which determines the effective collective linearized optomechanical interaction strengths [see Eq.~\rp{tildeg2-}]),
we can use Eq.~\rp{drivings-conditions-}, together with Eqs.~\rp{alphakm}, \rp{lambda} and \rp{XXYYalpha}, to determine the strengths and phases, $\ovl\epsilon_{k,m}$, of all the driving fields needed for the preparation of the cluster state~\rp{cluster-}. They are given by
\begin{eqnarray}
\ovl\epsilon_{k,j} &=& \frac{\wt g_k}{g_{k,j} }
\ppt{{\wt\omega_k-\omega_k-\Omega_j+i\frac{\kappa_k}{2}}}
\ppt{c_r\ \delta_{k,j}-\frac{i}{2}\ e^r\ \AAA_{k,j}}
\nn\\
\ovl\epsilon_{k,j+N}&=& \frac{\wt g_k}{ g_{k,j} }
\ppt{{\wt\omega_k-\omega_k+\Omega_j+i\frac{\kappa_k}{2}}}
\ppt{s_r\ \delta_{k,j}-\frac{i}{2}\ e^r\ \AAA_{k,j}}\ .
\end{eqnarray}
It is worth noting that the values of $r$ and $\wt g_k$ can be considered arbitrary only up to a certain extent. In fact, our model is applicable only when the relation in Eq.~\rp{RWA-} for the validity of the rotating wave approximation holds.
In the present case, this condition reduces to 
\begin{eqnarray}
\wt g_k\ \frac{g_{k,m}}{g_{k,m'}}\
\ppt{c_r\ \delta_{k,m'}+\frac{e^{r}}{2}\ \AAA_{k,m'}}
\ll
\abs{\Omega_m-\Omega_{m'}} 
\end{eqnarray}
for all $k,m,m'$ with $m\neq m'$ (see also \ref{condRWA}). 
In particular, if we assume that all the optomechanical couplings are equal, i.e. $g_{k,j}=g_{k',j'}$ for all $k,k',j,j'$, and $\wt g_k=\wt g_{k'}\equiv\wt g$ for all $k,k'$, and that the mechanical frequencies vary linearly with the mode index, i.e $\Omega_j=j\ \ovl\Omega$, then the rotating wave approximation is valid when 
\begin{eqnarray}\label{condRWA-}
\frac{\wt g\ e^r}{2\ \ovl\Omega}\ll 1\ .
\end{eqnarray}
This highlights the fact that both $\wt g$ and $r$ can not be taken too large.

%%%%%%%%%%%%%%%%%%%%%%%%%%%%%%%%%%%%%%%%%%%%%%%%%%%%%%%%%%%%%%%%%

\begin{figure}[t!]
\centering
\includegraphics[width=0.5\textwidth]{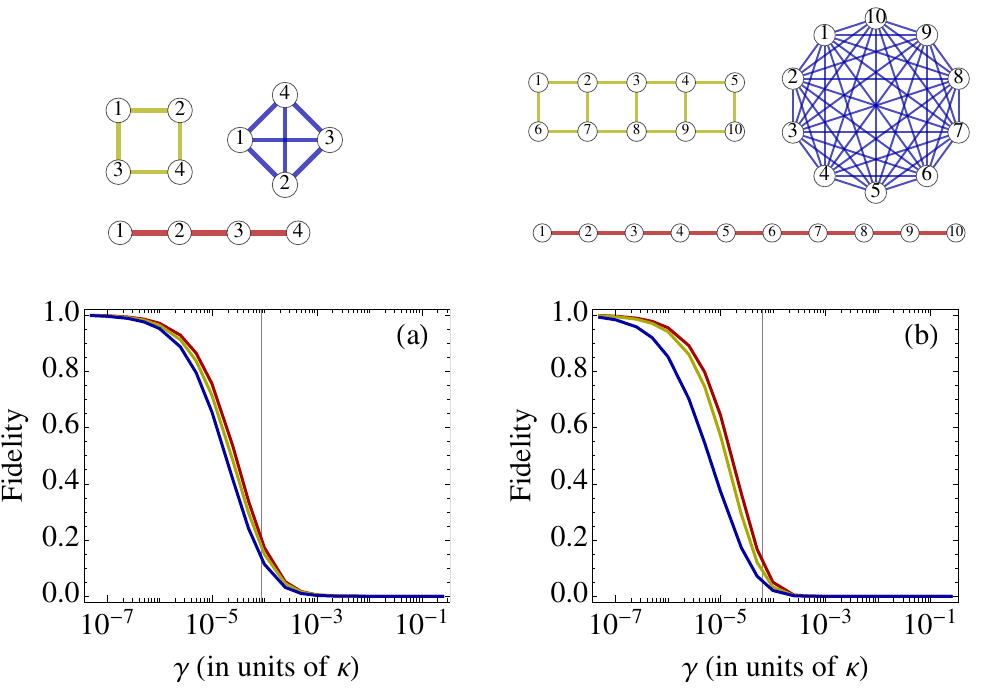}
\caption{Fidelity between the steady state of the engineered dynamics and the expected state, Eq.~\rp{Fid-}, as a function of the mechanical dissipation rate $\gamma\equiv\gamma_j$ for all $j$, for cluster states composed of (a) $N=4$ modes and (b) $N=10$ modes, and corresponding to linear (red, upper lines), rectangular (yellow, middle lines), and fully connected (blue, lower lines) graphs as depicted in the upper part of the figure. The nodes of the graphs represent the mechanical modes. 
The entry $\AAA_{j,j'}$ of the adjacency matrices [introduced in Eq.~\rp{C}] corresponding to these graphs is equal to $\AAA_{j,j'}=1$ if there is an edge connecting the nodes $j$ and $j'$, and equal to $\AAA_{j,j'}=0$ otherwise.
The other parameters are
$r=2$, temperature $=10$mK, $\Omega_j=j\,\ovl\Omega$ with $\ovl\Omega=2\pi\times10$MHz, $\kappa\equiv\kappa_k=0.02\ovl\Omega$, and $\wt g\equiv\wt g_k=0.16\kappa$ for all $k$. 
The vertical lines indicate the values for which $
\Xi^\star=4\ \wt g_k^2/\kappa_k
$ (quantum cooperativity for the fully connected graph equal to one) [see Eqs.~\rp{ineq0} and \rp{Xi}].
}
\label{fig2}
\end{figure}
\begin{figure*}[t!]
\centering
\includegraphics[width=0.9\textwidth]{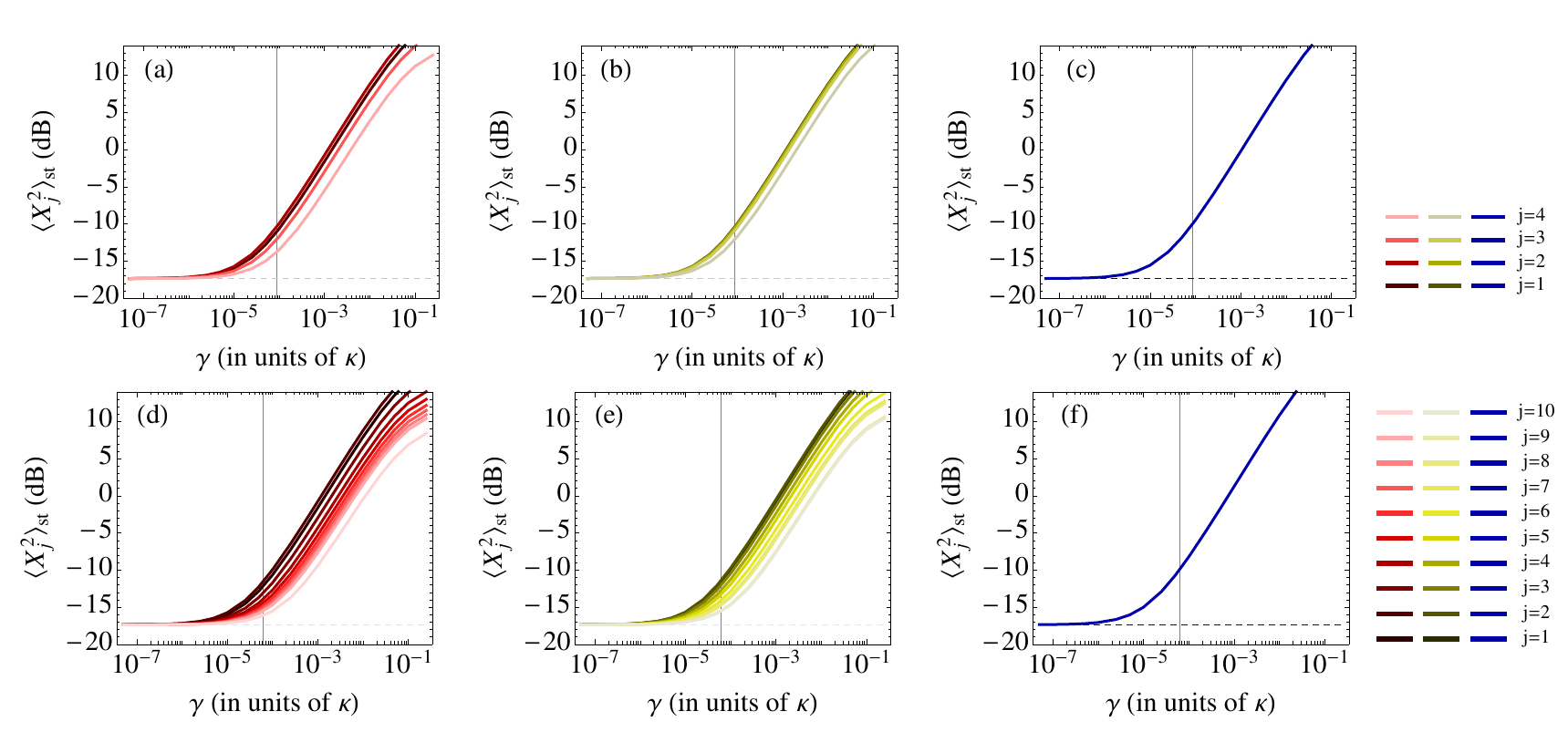}
\caption{
Variance of the nullifiers over the steady state 
$\langle X_j^2\rangle_{st}$ (solid lines), and corresponding value for the expected state (dashed lines), $\langle X_j^2\rangle_r$ [see Eqs.~\rp{X2r0} and \rp{X2st}] evaluated in dB, as a function of $\gamma$ and for the parameters of Fig.~\ref{fig2}. Panels (a), (b), and (c) are for $N=4$. Panels (d), (e) and (f) are for $N=10$. Red [(a) and (d)], yellow [(b) and (e)], and blue [(c) and (f)] lines are respectively for the linear, rectangular, and fully connected graphs reported in the upper part of Fig.~\ref{fig2}. Different shades of red and yellow indicate the results for different nullifiers [different $j$, see Eq.~\rp{nullifiers-}] as reported in the insets. In the case of the fully connected graphs [blue lines, (c) and (f)], the result is the same for all the nullifiers. 
The vertical lines indicate the values for which $\Xi^\star=4\ \wt g_k^2/\kappa_k
$ (quantum cooperativity for the fully connected graph equal to one) [see Eqs.~\rp{ineq0} and \rp{Xi}].  
 }
\label{fig3}
\end{figure*}

\begin{figure}[t!]
\centering
\includegraphics[width=0.5\textwidth]{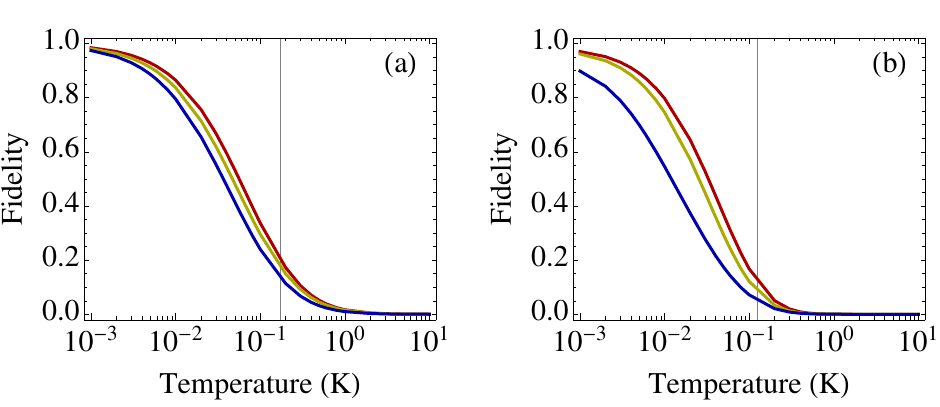}
\caption{
Fidelity, Eq.~\rp{Fid-}, as a function of the temperature. The other parameters and line styles are as in Fig.~\ref{fig2}, and $\gamma_j=5\times 10^{-6}\kappa$ for all $j$.
}
\label{fig4}
\end{figure}
\begin{figure*}[t!]
\centering
\includegraphics[width=0.9\textwidth]{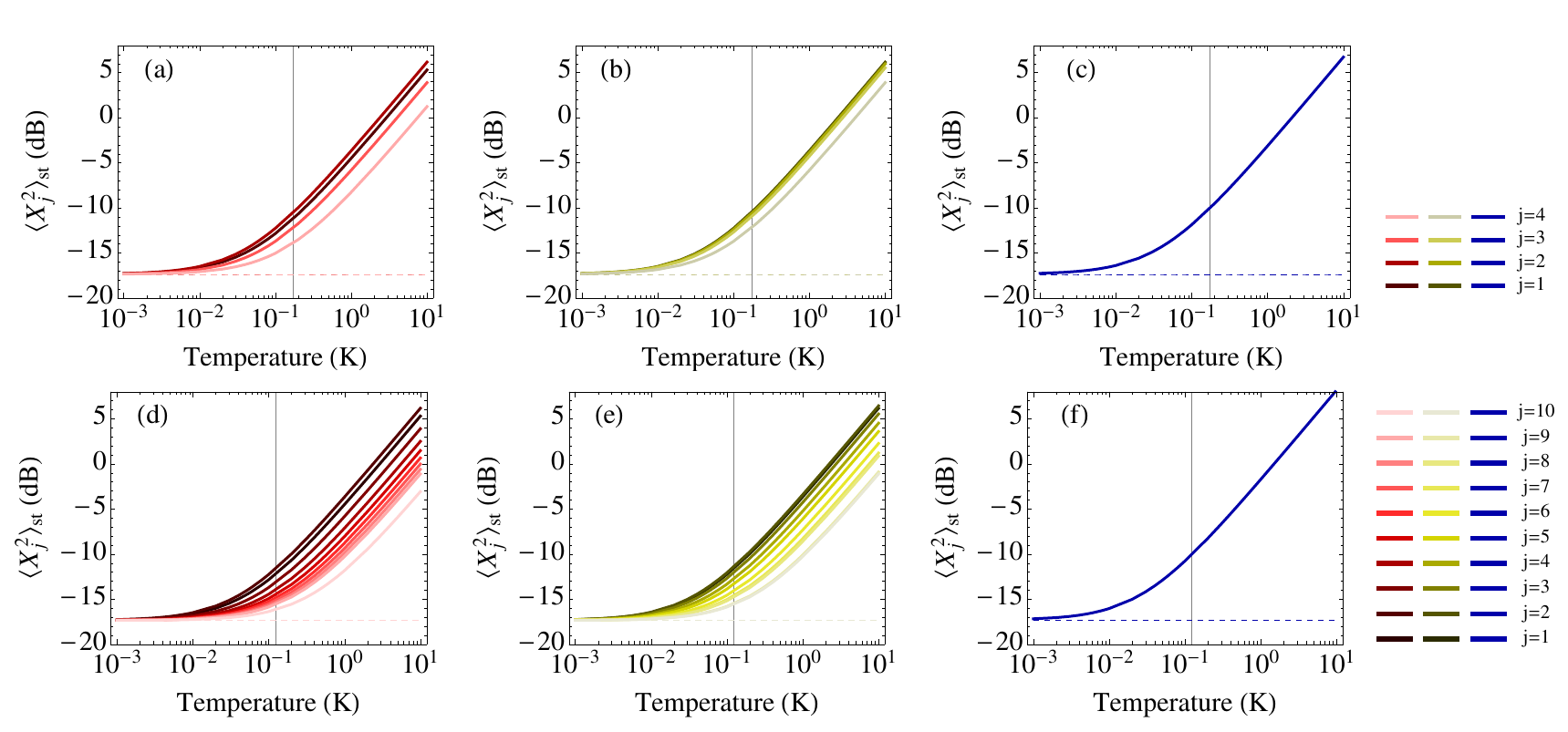}
\caption{
Variance of the nullifiers as a function of the temperature. The other parameters and line styles are as in Fig.~\ref{fig3}, and $\gamma_j=5\times 10^{-6}\kappa$ for all $j$.
}
\label{fig5}
\end{figure*}

\begin{figure}[t!]
\centering
\includegraphics[width=0.45\textwidth]{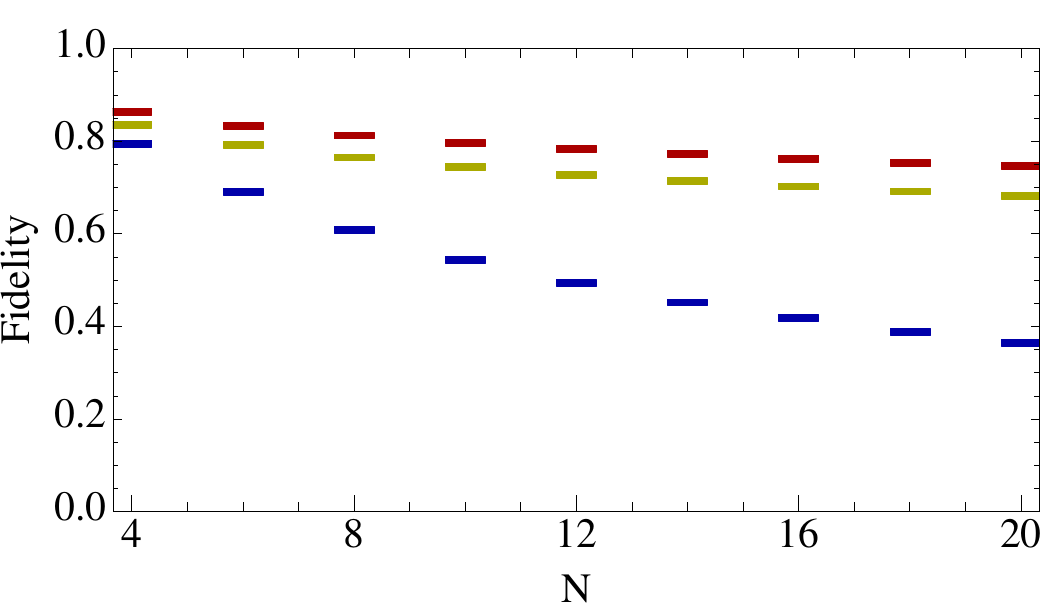}
\caption{
Fidelity, Eq.~\rp{Fid-}, as a function of the number of modes $N$. Upper (red), middle (yellow) and lower (blue) marks are, respectively, for the linear, rectangular and fully connected geometry. The other parameters and line styles are as in Fig.~\ref{fig2}, and $\gamma_j=5\times 10^{-6}\kappa$ for all $j$.
}
\label{fig6}
\end{figure}

\begin{figure}[t!]
\centering
\includegraphics[width=0.5\textwidth]{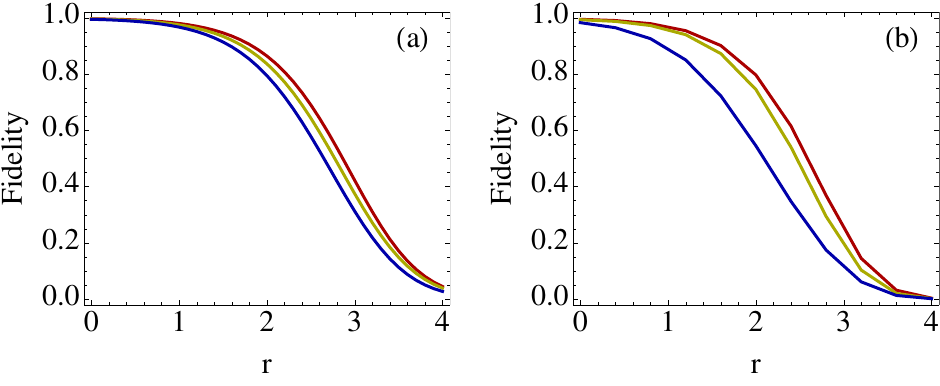}
\caption{
Fidelity, Eq.~\rp{Fid-}, as a function of $r$. The other parameters and line styles are as in Fig.~\ref{fig2}, and $\gamma_j=5\times 10^{-6}\kappa$ for all $j$.
}
\label{fig7}
\end{figure}
\begin{figure*}[t!]
\centering
\includegraphics[width=0.9\textwidth]{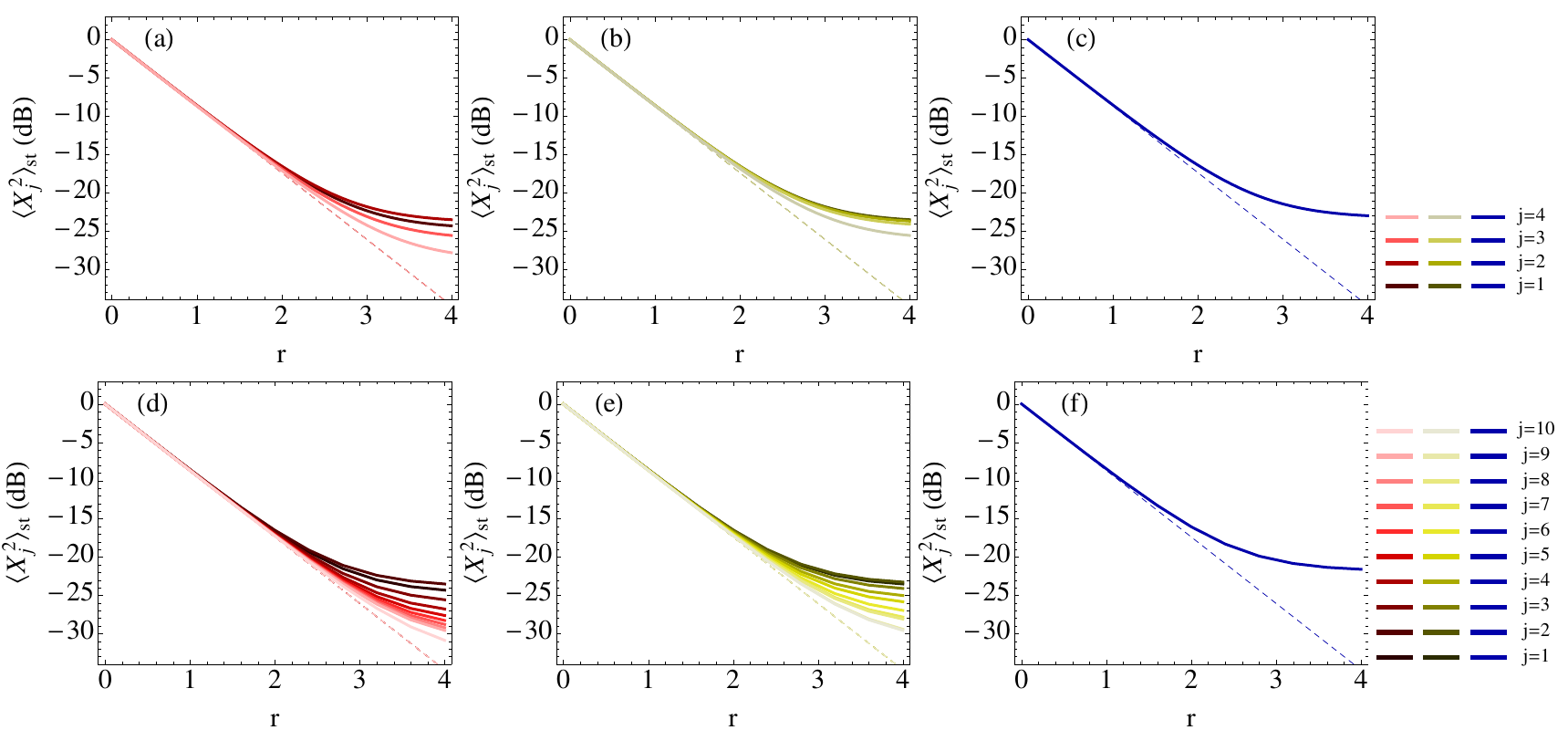}
\caption{
Variance of the nullifiers as a function of the squeezing parameter $r$. The other parameters and line styles are as in Fig.~\ref{fig3}, and $\gamma_j=5\times 10^{-6}\kappa$ for all $j$.
}
\label{fig8}
\end{figure*}

\begin{figure}[t!]
\centering
\includegraphics[width=0.5\textwidth]{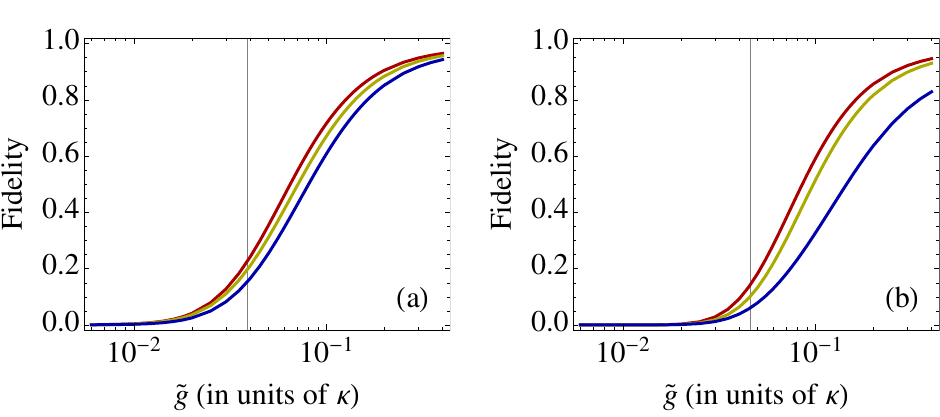}
\caption{
Fidelity, Eq.~\rp{Fid-}, as a function of $\wt g=\wt g_k$, for all $k$. The other parameters and line styles are as in Fig.~\ref{fig2}, and $\gamma_j=5\times 10^{-6}\kappa$ for all $j$.
}
\label{fig9}
\end{figure}

%%%%%%%%%%%%%%%%%%%%%%%%%%%%%%%%%%%%%%%%%%%%%%%%%%%%%

\section{Results}\label{Results}

Let us now quantify the performance of our protocol for the preparation of cluster states of various dimensions and geometries. We 
characterize the performance of our protocol in terms of the fidelity 
to prepare the state (see Sec.~\ref{Fidelity-} below) and in terms of the corresponding variance of the nullifiers (see Sec.~\ref{null-} below).

\subsection{Fidelity}\label{Fidelity-}

We analyze the fidelity between the expected state~\rp{cluster-} and the one obtained at the steady state of the quantum Langevin equations~\rp{QLEac-}. 
The fidelity can  be computed with the formula~\cite{spedalieri2013}
\begin{eqnarray}\label{Fid-}
F=\frac{2^N}{\sqrt{{\rm Det}\ppt{\VV_r\al{c}+\VV_{st}\al{c}}}}\ ,
\end{eqnarray}
where $\VV_r\al{c}$ and $\VV_{st}\al{c}$ 
are the covariance matrices of the position $x_j\al{c}=c_j+c_j\da$ and momentum $p_j\al{c}=-i\,c_j+i\,c_j\da$ operators for, respectively,  the expected states~\rp{cluster-} and the steady state of the system dynamics. 
In particular, 
when Eq.~\rp{CSU} is true, the correlations of the collective mechanical operators $c_j$, defined by Eq.~\rp{UbUc}, over the cluster state~\rp{cluster-}, are equal to Eq.~\rp{avcc}. And this entails that 
\begin{eqnarray}\label{Vrc}
\VV_r\al{c}=\id\ .
\end{eqnarray}
Moreover, the matrix $\VV_{st}\al{c}$ can be computed by solving the Eq.~\rp{QLEac-} using standard techniques as discussed in \ref{stst}.

\subsection{Variance of the nullifiers}\label{null-}

A cluster state can be equivalently defined as the zero eigenstate of the nullifiers 
\begin{eqnarray}\label{nullifiers-}
X_j=p_j-\sum_{j'=1}^N\ \AAA_{j,j'}\ x_{j'}
\end{eqnarray}
such that $X_j\ke{\psi_{cluster}}=0$ for all $j$. This entails that for realistic approximated cluster states, these operators are quantum squeezed (i.e., their variance is below the vacuum noise level). Consequently, the variance of the nullifiers constitutes a valid metric to characterize the quality of a cluster state. 
In particular, one finds that the variances of the nullifiers over the expected, approximated cluster state~\rp{cluster-} are
\begin{eqnarray}\label{X2r0}
\langle X_j^2\rangle_r&\equiv& \ 
\bbr{\psi_r}\ X_j^2\ \ke{\psi_r}=
\bbr{0}\ S_r\da\ C\da\ X_j\  X_{j}\ C\ S_r\ \ke{0}
\nn\\&=& \ \ee^{-2\,r}\ \ \ \ \ \ \forall\ j\ .
\end{eqnarray}
We note that this quantity is equal to the squeezing value of the original modes over which one applies the operator $C$~\rp{C} to define a cluster state [see Eq.~\rp{cluster-}]. In other terms, it is equal to the variance of the momentum quadratures $p_j$ of the original modes over the state $S_r\ke{0}$, i.e., 
$\bbr{0}S_r\da\ p_j^2\ S_r\ke{0}=\ee^{-2\,r}$. This quantity is used to estimate the effectiveness of a cluster state as a resource for measurement-based quantum computation. Specifically, various estimates~\cite{menicucci2014,fukui2018,walshe2019,fukui2023} indicate that a value of this quantity between -10 and -20 dB [that is, a value of $10\,\log_{10}\ppt{\ee^{-2\,r}}$ between -10 and -20] is needed to achieve fault-tolerant measurement-based quantum computation.
We also note that Ref.~\cite{walshe2019} suggests that the value of squeezing $\bbr{0}S_r\da\ p_j^2\ S_r\ke{0}$ determines the potentiality of a cluster state regardless of the purity of the state itself.
Therefore, an approximate cluster state can be a useful resource for quantum computation even if the fidelity with a pure, ideal cluster state is particularly low, provided that the squeezing is sufficiently large.
For this reason, the value of the variance of the nullifiers $\av{X_j^2}_{st}$ over the steady state represents a useful estimate of the quality of the steady state as a valid resource for quantum computation.
It can be expressed as follows. 
We consider the vector of operators $\vx=\ppt{x_1,\dots x_N,p_1,\dots p_N}^T$
and the $N\times 2N$  matrix 
\begin{eqnarray}
\OO
=\ppt{\mmat{cc}{
-\AAA &  \  \id
}}\ ,
\end{eqnarray}
such that the nullifiers can be written as $X_j=\pg{\OO\ \vx}_j$.
Correspondingly, $\av{X_j^2}_{st}$ can be expressed, in terms of the matrix of correlations  associated to the original modes $\VV_r\al{b}$ [the element of which are $\pg{\VV_{st}\al{b}}_{\ell,\ell'}=\ppt{\av{\vx_\ell\ \vx_{\ell'}}_{st}+\av{\vx_{\ell'}\ \vx_\ell}_{st}}/2$], as the diagonal elements of the matrix $\OO\ \VV_{st}\al{b}\ \OO^T$, i.e. 
\begin{eqnarray}\label{X2st}
\langle X_j^2 \rangle_{st}=\pg{ \OO\ \VV_{st}\al{b}\ \OO^T }_{j,j}\ .
\end{eqnarray}

\subsection{Numerical results}

We compute numerically the covariance matrix $\VV_{st}\al{c}$, of the collective modes and the corresponding matrix $\VV_{st}\al{b}$ for the original modes, as detailed in \rp{stst}.
These matrices are used to evaluate the fidelity using Eqs.~\rp{Fid-} and \rp{Vrc} and the variance of the nullifiers using Eq.~\rp{X2st}. 
We show the results for three different graph geometries, linear (red lines), rectangular (yellow lines) and fully connected (blue lines) and various number of modes, as depicted in the upper part of Fig.~\ref{fig2}.

First, we examine the effect of noise. The system reaches the expected state~\rp{cluster-} exactly, only if the mechanical dissipation rates $\gamma_j$ are negligible (see Figs.~\ref{fig2} and \ref{fig3}). At finite $\gamma_j$ the final state is closer to the expected result the smaller the temperature, as depicted in Figs.~\ref{fig4} and \ref{fig5}. It's important to note that in general, even at zero temperature, quantum noise prevents the exact generation of the expected state. However, sufficiently large mechanical quality factors, as those used in Figs.~\ref{fig4}-\ref{fig9}, allow for very good cluster state preparation at low temperatures.
Furthermore, the detrimental effect of noise becomes more pronounced with a larger number of modes and for cluster states described by graphs with larger number of connections.
In particular, we analyze the scaling with system size in Fig.~\ref{fig6}. We observe that the preparation is not very sensitive to the number of modes itself but depends more strongly on the number of connections of each mode. This is evident comparing the results of the linear and rectangular clusters with those of the fully connected one, which exhibit much lower fidelity at large $N$.

In general, we observe good state preparation when the light-induced mechanical dissipation rate, which is of the order of $4\,\wt g_k^2/\kappa_k$~\cite{aspelmeyer2014}, is sufficiently large to overcome the effect of thermal noise, which affects the dynamics of the collective modes~\rp{cj-}. Here, thermal noise is characterized by the correlations of the collective noise operators $\av{f_j\da(t)\,f_j(t')}=\delta(t-t')\,\Xi_j$, where [see Eqs.~\rp{fk-}, \rp{drivings-conditions-}, \rp{avff}-\rp{Xik}] 
\begin{eqnarray}\label{Xij}
\Xi_j\sim& &\ \frac{e^{2r}}{2}
\pq{
\gamma_j\ppt{n_j+\frac{1}{2}}+
\sum_{j'\in \pg{N_j}}\gamma_{j'}\ppt{n_{j'}+\frac{1}{2}}
}
\ ,
\end{eqnarray}
and where $\pg{N_j}$, in the subscript of the summation, indicates the set of modes adjacent to the mode $b_j$ according to the matrix $\AAA$.
Therefore, the condition for a faithful preparation of a cluster state is
\begin{eqnarray}\label{ineq0}
\frac{4\ \wt g_k^2}{\kappa_k\ \Xi_j}\gg1\ ,
\end{eqnarray} 
with $\Xi_j$ defined in Eq.~\rp{Xij}. 
This quantity is the effective or quantum cooperativity~\cite{ren2020a} for the collective mechanical Bogoliubov mode $c_j$~\rp{cj-}.
We remark that it depends not only on the natural mechanical thermal noise, but also on the cluster state we aim to achieve through the factor $e^{2 r}$ and the sets of adjacent modes $\pg{N_j}$.
We also note that, in the case of a fully connected cluster, $\Xi_j$ is independent of $j$ and is given by
\begin{eqnarray}\label{Xi}
\Xi^\star=
\frac{e^{2\,r}\ }{2}\sum_{j'=1}^N\gamma_{j'}\ \ppt{n_{j'}+\frac{1}{2}}\ ,
\end{eqnarray}
for all $j$.
In Figs.~\ref{fig2}-\ref{fig5} and \ref{fig9}, we emphasize the significance of the condition in Eq.~\rp{ineq0}, by adding vertical lines corresponding to the values where the quantum cooperativity of a fully connected graph is one, that is where $
\Xi^\star=4\ \wt g_k^2/\kappa_k$.
For these values, the fidelity starts to increase to noticeable amounts, but, at the same time, the variance of the nullifiers is already very low, indicating strong squeezing. 
Specifically, this value is at the level of recent estimates~\cite{fukui2018,fukui2023} for the thresholds of squeezing required for fault-tolerant quantum computation. 
Therefore, although not exactly equal to the expected state, the steady state is already a good approximation of a cluster state. These results are evaluated for mechanical frequencies in the range $2\pi\times\,10-2\pi\times\,100$MHz, in the resolved sideband regime, and at a temperature of 10mK (except for the results shown as a function of temperature itself in Figs.~\rp{fig4} and \ref{fig5}). The corresponding mechanical quality factors, in Figs.~\ref{fig4}-\ref{fig9}, range between $\Omega_1/\gamma_1=10^7$ and $\Omega_{20}/\gamma_{20}=2\times 10^8$. However, Fig.~\ref{fig3} show that the variance of the nullifiers remains quite low also for quality factors roughly one order of magnitude smaller, in the range $10^6-10^7$.
These values are sufficient to achieve good
preparation of cluster states with $r=2$. For larger values of $r$ the fidelity decreases (see Fig.~\ref{fig7})
because, as discussed above, the effective noise of the Bogoliubov modes increases [see Eqs.~\rp{Xij} and \rp{ineq0}]. Nevertheless, at the same time the variance of the nullifiers remains very low (see Fig.~\ref{fig8}), indicating that the steady state is still a good approximation of a cluster state.

To achieve higher fidelities one should either reduce the value of squeezing $r$ or select larger mechanical frequencies and/or larger values of $\wt g$.
However, increasing both $\wt g$ and $r$ simultaneously may be problematic due to the requirements
for the validity of the rotating wave approximation expressed in Eq.~\rp{condRWA-}. 
This condition remains valid for all the values of $r$ reported in Figs.~\ref{fig7} and \ref{fig8} (where for $r=4$, we find $\frac{\wt g\ e^r}{2\ \ovl\Omega}\sim 0.09$), and for all the values of $\wt g$ used in Fig.~\rp{fig9} (which is evaluated for $r=2$). However, for larger value of $r$ and $\wt g$, the validity of the rotating wave approximation can be questionable unless one increases also the mechanical frequencies (or, more precisely, the value of ${\rm min}_{j\neq j'\in\pg{1,\dots N}}\pq{\Omega_j-\Omega_{j'}}$).

\subsection{Experimental implementation}

We comment here on the possibility of observing and applying these results in real experiments. The parameters used for the numerical simulation, though challenging, are within present-day technological capabilities in both optical and microwave regimes. Mechanical resonator frequencies in optomechanics can vary widely, from tens of kHz to tens of GHz~\cite{aspelmeyer2014}. Notably, various multimode optomechanical experiments have reported mechanical frequencies in the tens of MHz, similar to those used in our simulations~\cite{
massel2012,fan2015,bernier2017a,bernier2017a,peterson2017a,ockeloen-korppi2018,
ruesink2018,colombano2019,kotler2021,mercierdelepinay2021,delpino2022}. While the quality factors in our simulation are quite large, it's worth noting that the largest reported mechanical quality factor to date is $>10^9$~\cite{seis2022,planz2023}, which is one to three orders of magnitude larger than those used here.

Additionally, our numerical results correspond to cryogenic temperatures. The temperature of 10 mK used in Figs.~\ref{fig2}, \ref{fig3}, \ref{fig6}-\ref{fig9} is demanding, but we observe that in Fig.~\ref{fig5}, a high level of squeezing (and thus good cluster state preparation) is achieved even at temperatures around 100 mK. Moreover, the temperature requirement can be further relaxed by employing higher mechanical frequencies in the GHz range~\cite{riedinger2018,fiaschi2021,kharel2022,youssefi2022a,mercade2023}.

The generated state can be probed by extending the approach discussed in Ref.~\cite{li2015}. One should utilize $N$ additional cavity modes and drive them with additional probe fields resonant with the red mechanical sidebands to transfer the mechanical state to the electromagnetic field. The reflected light can then be measured by homodyne detection, and the mechanical covariance matrix can be reconstructed. A similar detection strategy can be employed to perform the measurements needed for a quantum algorithm when using this state for measurement-based quantum computation.

\section{Conclusions}\label{Conclusions}

In conclusions, we have studied a scheme for the dissipative stabilization of the quantum state of a multimode optomechanical system, constituted of $N$ mechanical and $N$ optical modes. 
We have identified the conditions that the driving fields must satisfy in order to achieve full control over the quantum properties of the collective steady state of $N$ mechanical resonators.
We have demonstrated that this can be attained for mechanical resonators of different frequencies in the resolved sideband regime, by employing driving fields with $2N^2$ frequency components that are resonant with all the blue and red sidebands of all the optical modes, corresponding to all the mechanical modes.
In particular, we have applied this result to the stabilization of mechanical cluster states and we have identified the corresponding conditions for the strengths of all the driving components.
We have shown that we achieve good preparation, in terms of the fidelity and of the variance of the nullifiers, using system parameters within present day technological capabilities.

%%%%%%%%%%%%%%%%%%%%%%%%%%%%%%%%%%%%%%%%%%%%%%%%%%%%%%%%%%%%%%
%\begin{acknowledgments}
\ack
We acknowledge financial support from NQSTI within PNRR MUR project PE0000023-NQSTI.
%\end{acknowledgments}

%%%%%%%%%%%%%%%%%%%%%%%%%%%%%%%%%%%%%%%%%%%%%%%%%%%%%%%%%%%%%%
\appendix
\setcounter{section}{0}
%%%%%%%%%%%%%%%%%%%%%%%%%%%%%%%%%%%%%%%%%%%%%%%%%%%%%%%%%%%%%%

\section{Derivation of the approximated model}\label{Model}

The full system is described in terms of the bosonic annihilation and creation operators $B_j$ and $B_j\da$ of the mechanical modes and 
$A_k$ and $A_k\da$ of the optical ones.
Correspondingly, the full Hamiltonian is 
\begin{eqnarray}
H &=& \sum_{k=1}^{N} \omega_k\ A_k^\dagger A_k + \sum_{j=1}^{N} \Omega_j\ B_j B_j^\dagger 
%\\& &
+ \sum_{k,j=1}^{N} g_{kj}\ A_k^\dagger A_k \ppt{B_j + B_j^\dagger} 
\nn\\& &
+ \sum_{k=1}^{N} \epsilon_k(t)\ A_k^\dagger + \epsilon_k^*(t)\ A_k\ .
\end{eqnarray}
Including dissipation and noise of the optical and mechanical modes at rate $\kappa_k$ and $\gamma_j$ respectively, we can write the corresponding quantum Langevin Equations
\begin{eqnarray}
\dot{A}_{k}&=& i[H,A_{k}]-\frac{\kappa_{k}}{2}\,A_{k}+\sqrt{\kappa_{k}}\ a_{k}^{in}
\nn\\& =&
-\ppt{i\omega_{k}+\frac{\kappa_{k}}{2}}A_{k}
-i\epsilon_{k}(t)-i\sum_{j=1}^{N}g_{kj}\ppt{B_{j}+B_{j}^{\dagger}}A_{k}+\sqrt{\kappa_{k}}\ a_{k}^{in}
\nn\\
\dot{B}_{j}	&=& i[H,B_{j}]-\frac{\gamma_{j}}{2}B_{j}+\sqrt{\gamma_{j}}\ b_{j}^{in}
\nn\\
&=& -\ppt{i\Omega_{j}+\frac{\gamma_{j}}{2}}B_{j}-i\sum_{k=1}^{N}g_{kj}\ A_{k}^{\dagger}A_{k}+\sqrt{\gamma_{j}}\ b_{j}^{in}
\end{eqnarray}
where $a_k^{in}$ and $b_j^{in}$ are the zero-average input noise operators 
introduced in the main text.

\subsection{Linearization and expansion for small optomechanical couplings}\label{Linearization}

As usual with optomechanical systems we consider the fluctuations of the system variables around the corresponding average values $\av{A_k}=\alpha_k(t)$ and $\av{B_j}=\beta_j(t)$ which fulfill the equations
\begin{eqnarray}
\dot{\alpha}_k(t) &=&  -\ppt{i\omega_k+\frac{\kappa_k}{2}}\alpha_k(t) - i\epsilon_k(t) 
%\nn\\& &
- i\sum_{j=1}^{N}g_{kj} \pq{\beta_j(t)+\beta_j(t)^*}\alpha_k(t) 
\nn\\
\dot{\beta}_j(t) &=&  -\ppt{i\Omega_j+\frac{\gamma_j}{2}}\beta_j(t) - i\sum_{k=1}^{N}g_{kj}\ \abs{\alpha_k(t)}^2\ .
\end{eqnarray}
These equations can be solved and their solution can be used in the equation for the fluctuations that we report below. Specifically, we compute the steady state, expanding it in powers of the coupling coefficients $g_{kj}$. Here, we assume that these coefficients are sufficiently small, allowing us to consider only the lowest order terms, that is the zeroth order for $\alpha_k$ and the first order for $\beta_j$. 
We find
\begin{eqnarray}
\alpha_k(t)&=& \
e^{-i\ppt{\omega_k-i\frac{\kappa_k}{2}}t}\alpha_k(0)+\sum_{m=1}^{2\,N}\ \ovl\alpha_{km}\pq{ e^{-i\,\lambda_{km}\ t}-e^{-i\ppt{\omega_k-i\frac{\kappa_k}{2}}t} }
%\nn\\& &\ 
+o\ppt{g_{kj}} 
\nn\\&=& \
\sum_{m=0}^{2\,N}\ \ovl\alpha_{km}\ e^{-i\,\lambda_{km}\ t}+o\ppt{g_{kj}} 
\end{eqnarray}
where we have introduced the definitions
\begin{eqnarray}
\lambda_{k,0}&\equiv& \ \omega_k-i\frac{\kappa_k}{2}
\end{eqnarray}
and
\begin{eqnarray}\label{baralpha}
\ovl \alpha_{km}&=& \ \frac{\ovl\epsilon_{km}}{\lambda_{km}-\omega_k+i\frac{\kappa_k}{2}}  \ \ \ \ \ \ {\rm for}\ m>0
\\
\ovl \alpha_{k,0}&=& \ \alpha_k(0)-\sum_{m=1}^{2\,N}\ \ovl \alpha_{km}
\end{eqnarray}
and correspondingly
\begin{eqnarray}
\beta_j(t)&=& \
e^{-i\ppt{\Omega_j-i\frac{\gamma_j}{2}}t}\beta_j(0)
%\nn\\& &\
+\sum_{k=1}^N\sum_{m,m'=0}^{2\,N}\ \ovl\beta_{jkmm'}\pq{ e^{-i\ppt{\lambda_{km}-\lambda_{km'}^*} t}-e^{-i\ppt{\Omega_j-i\frac{\gamma_j}{2}}t} }
%\nn\\& &
+o\ppt{g_{kj}} 
\nn\\&=& \
\sum_{k=0}^N\sum_{m,m'=0}^{2\,N}\ \ovl\beta_{jkmm'}\ e^{-i\,\xi_{jkmm'}\ t}+o\ppt{g_{kj}} 
\end{eqnarray}
where
\begin{eqnarray}
\xi_{j0mm'}&=& \ \Omega_j-i\frac{\gamma_j}{2}
\nn\\
\xi_{jkmm'}&=& \ \lambda_{km}-\lambda_{km'}^*
\nn\\
\ovl \beta_{jkmm'}&=& \ \frac{g_{kj}\ \ovl\alpha_{km}\ \ovl\alpha_{km'}^*}{\lambda_{km}-\lambda_{km'}^*-\Omega_j+i\frac{\gamma_j}{2}} \ \ \ \ \  {\rm for}\ k>0
\nn\\
\ovl \beta_{j000}&=& \  \beta_j(0)-\sum_{k=1}^{N}\sum_{m,m'=0}^{2\,N}\ \ovl\beta_{jkmm'}
\nn\\
\ovl \beta_{j0mm'}&=& \ 0 \ \ \ \ \  {\rm for}\ m,m'>0\ .
\end{eqnarray}
In particular in the long time limit $t\to\infty$ we find
\begin{eqnarray}
\alpha_k(t)&\simeq&\
\sum_{m=1}^{2\,N}\ \ovl\alpha_{km}\ e^{-i\,\lambda_{km}\ t}
+o\ppt{g_{kj}^2} 
\nn\\
\beta_j(t)&\simeq&\
\sum_{k=1}^N\sum_{m,m'=1}^{2\,N}\ \ovl\beta_{jkmm'}\ e^{-i\ppt{\lambda_{km}-\lambda_{km'}}t}
+o\ppt{g_{kj}^3} 
\nn\\&=& \
\beta_j\oo+\beta_j'(t)
+o\ppt{g_{kj}^3} 
\end{eqnarray}
where $\beta_j\oo$ accounts for the time independent part of $\beta_j(t)$, that is
\begin{eqnarray}
\beta_j\oo&=& \ \sum_{k=1}^N\sum_{m=1}^{2\,N}\ \ovl\beta_{jkmm}\ .
\end{eqnarray}

The equations for the fluctuations
\begin{eqnarray}
\wt a_k&=& \ A_k-\alpha(t)
\nn\\
\wt b_j&=& \ B_j-\beta_j(t)
\end{eqnarray}
can be linearized assuming sufficiently strong drivings. Specifically, introducing the shifted optical frequencies $\wt\omega_k=\omega_k+\delta_k$ with
\begin{eqnarray}\label{deltak}
\delta_k=- 2\,i\sum_{j=1}^{N}g_{kj}\ {\rm Re}\pq{\beta_j\oo}\ ,
\end{eqnarray}
one finds the following equations for the fluctuations 
\begin{eqnarray}
\dot{\wt a}_{k} &=&  -\ppt{i\ \wt\omega_k+\frac{\kappa_k}{2}}\wt a_k - 2\,i\sum_{j=1}^{N}g_{kj}\ {\rm Re}\pq{\beta_j'(t)}\wt a_k 
\nn\\& &
- i\sum_{j=1}^{N}g_{kj}\ppt{\wt b_j+\wt b_j^\dagger}\alpha_k(t) + \sqrt{\kappa_k}\ a_k^{in} 
\nn\\
\dot{\wt b}_{j} &=&  -\ppt{i\Omega_j+\frac{\gamma_j}{2}}\wt b_j - i\sum_{k=1}^{N}g_{kj}\pq{\wt a_k^\dagger\ \alpha_k(t)+\wt a_k\ \alpha_k(t)^*} 
%\nn\\& &
+ \sqrt{\gamma_j}\ b_j^{in} \ .
\end{eqnarray}
In the interaction picture we find the equations for the field operators in a rotating frame $a_k$ and $b_j$, defined by $\wt a_k=e^{-i\,\wt\omega_k\ t}\,a_k$ and $\wt b_j=e^{-i\,\Omega_j\ t}\,b_j$,
\begin{eqnarray}
\dot{a}_k &=& \ -\frac{\kappa_k}{2}\,a_k 
- 2\,i\sum_{j=1}^{N}g_{kj}\ {\rm Re}\pq{\beta_j'(t)}a_k 
- i\sum_{j=1}^{N}\sum_{m=1}^{2\,N}\ 
g_{kj}\ \ovl\alpha_{km}
\nn\\& &\ \times
\pq{
b_j\ e^{i\ppt{\wt\omega_k-\Omega_j-\lambda_{km}} t}+
b_j^\dagger\ e^{i\ppt{\wt\omega_k+\Omega_j-\lambda_{km}} t}
} 
- \sqrt{\kappa_k}\ a_k^{in} 
\nn\\
\dot{b}_{j} &=&  -\frac{\gamma_j}{2}\,\wt b_j - i\sum_{k=1}^{N}\sum_{m=1}^{2\,N}\ 
g_{kj}
%\nn\\& &\ \times
\pq{
\ovl\alpha_{km}\ a_k^\dagger\ e^{i\ppt{\wt\omega_k+\Omega_j-\lambda_{km}} t} +
\ovl\alpha_{km}^*\ a_k\ e^{-i\ppt{\wt\omega_k-\Omega_j-\lambda_{km}} t}
} 
\nn\\& &
- \sqrt{\gamma_j}\ b_j^{in} \ .
\end{eqnarray}
In Eq.~\rp{dotadotb} we have not reported, for simplicity, the term proportional to $\beta'_j$ which, as shown below, is negligible.

\subsection{Full conditions for the rotating wave approximation}
\label{condRWA}

Let us now consider the conditions for the driving frequencies in Eq.~\rp{lambda} such that, for $m,m'\leq N$,
\begin{eqnarray}
\ovl \alpha_{k,m}&=& \ \frac{\ovl\epsilon_{k,m}}{\delta_k-\Omega_m+i\frac{\kappa_k}{2}} 
\nn\\
\ovl \alpha_{k,m+N}&=& \ \frac{\ovl\epsilon_{k,m+N}}{\delta_k+\Omega_m+i\frac{\kappa_k}{2}} 
\nn\\
\ovl \beta_{j,k,m,m'}&=& \ \frac{g_{k,j}\ \ovl\alpha_{k,m}\ \ovl\alpha_{k,m'}^*}{\Omega_{m'}-\Omega_m-\Omega_j+i\frac{\gamma_j}{2}} 
\nn\\
\ovl \beta_{j,k,m,m'+N}&=& \ \frac{g_{k,j}\ \ovl\alpha_{k,m}\ \ovl\alpha_{k,m'+N}^*}{-\Omega_m-\Omega_{m'}-\Omega_j+i\frac{\gamma_j}{2}} 
\nn\\
\ovl \beta_{j,k,m+N,m'}&=& \ \frac{g_{k,j}\ \ovl\alpha_{k,m+N}\ \ovl\alpha_{k,m'}^*}{\Omega_m+\Omega_{m'}-\Omega_j+i\frac{\gamma_j}{2}} 
\nn\\
\ovl \beta_{j,k,m+N,m'+N}&=& \ \frac{g_{k,j}\ \ovl\alpha_{k,m+N}\ \ovl\alpha_{k,m'+N}^*}{\Omega_m-\Omega_{m'}-\Omega_j+i\frac{\gamma_j}{2}} \ .
\end{eqnarray}
The full set of conditions for the validity of the rotating wave approximation (including also those corresponding to the average mechanical fields $\beta_{j,k,m,m'}$) is, 
\begin{eqnarray}\label{RWA}
\abs{g_{k,j}\ \ovl\alpha_{k,m}}, 
\abs{g_{k,j}\ \ovl\alpha_{k,m+N}}&\ll&\ \abs{\Omega_j-\Omega_m} 
\ \ \ \ \   {\rm for}\ j\neq\ m 
\nn\\
\abs{2\,g_{k,j}\ \ovl\beta_{j,k,m,m'}},
\abs{2\,g_{k,j}\ \ovl\beta_{j,k,m+N,m'+N}}&\ll&\ \abs{\Omega_m-\Omega_{m'}}  
\ \ \ \ \   {\rm for}\ m\neq\ m' 
\nn\\
\abs{2\,g_{k,j}\ \ovl\beta_{j,k,m,m'+N}},
\abs{2\,g_{k,j}\ \ovl\beta_{j,k,m+N,m'}}&\ll&\ \abs{\Omega_m+\Omega_{m'}}  
\ .
\end{eqnarray}
In the main text we have indicated only the first one, 
which, if fulfilled, implies the validity of also the other two conditions.

Using Eqs.~\rp{XXYYalpha} and \rp{drivings-conditions-}, we find that the conditions for the validity of the rotating wave approximation take the form 
%(for $m\neq m'$)
\begin{eqnarray}
\abs{\Omega_m-\Omega_{m'}}\gg& &\  
\wt g_k\ \frac{g_{k,m}}{g_{k,m'}}\
\ppt{c_r\ \delta_{k,m'}+\frac{e^{r}}{2}\ \AAA_{k,m'}} 
%\ \ \ \ {\rm for\ all}\ k
\end{eqnarray}
and 
%(for $m\neq m'$)
\begin{eqnarray}\label{RWA2}
\abs{\Omega_m-\Omega_{m'}}\gg& &\ 
\frac{
2\ \wt g_k^2\ g_{k,j}^2
\
\ppt{c_r\ \delta_{k,m}+\frac{e^r}{2}\AAA_{k,m}}
\ppt{c_r\ \delta_{k,m'}+\frac{e^r}{2}\AAA_{k,m'}}
}{
g_{k,m}\ g_{k,m'}
\sqrt{\min\pg{\ppt{\Omega_{m'}-\Omega_m\pm\Omega_j}^2}+\frac{\gamma_j^2}{4}}
}
\nn\\
\abs{\Omega_m+\Omega_{m'}}\gg& &\ 
\frac{
2\ \wt g_k^2\ g_{k,j}^2
\
\ppt{c_r\ \delta_{k,m}+\frac{e^r}{2}\AAA_{k,m}}
\ppt{c_r\ \delta_{k,m'}+\frac{e^r}{2}\AAA_{k,m'}}
}{
g_{k,m}\ g_{k,m'}
\sqrt{\ppt{\Omega_{m'}+\Omega_m-\Omega_j}^2+\frac{\gamma_j^2}{4}}
}\ .   
\end{eqnarray}

\subsection{The collective mechanical noise}
\label{collnoise}

Under conditions~\rp{RWA}-\rp{RWA2} we find the approximated quantum Langevin equations~\rp{Eqab} and \rp{QLEac-}, where
the collective mechanical noise operators $f_k$~\rp{fk-} are characterized by the correlation functions
\begin{eqnarray}\label{avff}
\av{f_k(t)\ f_{k'}\da(t')}&=& \ \delta(t-t')\ \Xi_{k,k'}\al{-+}
\nn\\
\av{f_k\da(t)\ f_{k'}(t')}&=& \ \delta(t-t')\ \Xi_{k,k'}\al{+-}
\nn\\
\av{f_k(t)\ f_{k'}(t')}&=& \ \delta(t-t')\ \Xi_{k,k'}\al{--}
\nn\\
\av{f_k\da(t)\ f_{k'}\da(t')}&=& \ \delta(t-t')\ \Xi_{k,k'}\al{++}
\end{eqnarray}
with %[see Eq.~\rp{drivings-conditions-}]
\begin{eqnarray}\label{FF}
\Xi_{k,k'}\al{-+}&=& \ \sum_{j=1}^N\ \gamma_j
\pq{
\XX_{k,j}\ \XX_{k',j}^*\ \ppt{n_j+1}+
\YY_{k,j}\ \YY_{k',j}^*\ n_j
}
\nn\\
\Xi_{k,k'}\al{+-}&=& \ \sum_{j=1}^N\ \gamma_j
\pq{
\XX_{k,j}^*\ \XX_{k',j}\ n_j+
\YY_{k,j}^*\ \YY_{k',j}\ \ppt{n_j+1}
}
\nn\\
\Xi_{k,k'}\al{--}&=& \ \sum_{j=1}^N\ \gamma_j
\pq{
\XX_{k,j}\ \YY_{k',j}\ \ppt{n_j+1}+
\YY_{k,j}\ \XX_{k',j}\ n_j
}
\nn\\
\Xi_{k,k'}\al{++}&=& \ \sum_{j=1}^N\ \gamma_j
\pq{
\XX_{k,j}^*\ \YY_{k',j}^*\ n_j+
\YY_{k,j}^*\ \XX_{k',j}^*\ \ppt{n_j+1}
}\ .
\end{eqnarray}
Note that in the main text, in Eq.~\rp{Xij}, we have used the symbol $\Xi_k$ in place of $\Xi\al{+-}_{k,k}$.
In particular, using the expressions of the matrices $\XX$ and $\YY$~\rp{drivings-conditions-}, that correspond to a cluster state, we find
\begin{eqnarray}\label{Xik}
\Xi_k&\equiv& \ \Xi\al{+-}_{k,k}
\\
&=& \ 
c_r^2\ \gamma_k\ n_k+\frac{e^{2r}\,}{4} \sum_{j=1}^N \gamma_j\ \AAA_{k,j}^2\ n_j 
%\nn\\& &\ 
+
s_r^2\ \gamma_k\ (n_k+1)+\frac{e^{2r}\,}{4} \sum_{j=1}^N \gamma_j\ \AAA_{k,j}^2\ (n_j+1) 
\nn
\end{eqnarray}
which approaches Eq.~\rp{Xij} when $e^{2\,r}\gg 1$.

\section{The steady state}
\label{stst}

In order to analyze the steady state correlations it is useful to express Eq.~\rp{QLEac-} in matrix form in terms of the vector of operators 
$\va=\ppt{a_1,\dots,a_N,c_1,\dots,c_N,a_1\da,\dots,a_N\da,c_1\da,\dots,c_N\da}^T$, as
\begin{eqnarray}
\dot\va=\MM\ \va+\vf
\end{eqnarray}
with $\MM$ given by the block matrix
\begin{eqnarray}
\MM=\ppt{\mmat{cccc}{
-\KK/2    &  -i\ \GG &  {\it 0} &  {\it 0}
\\
-i\ \GG & -\WW    &  {\it 0} & -\TT
\\
{\it 0} & {\it 0} & -\KK/2  & i\ \GG
\\
{\it 0} & -\TT^*  & i\ \GG  & -\WW^*
}}
\end{eqnarray}
where $\KK$ and $\GG$ are diagonal matrices with elements $\KK_{k,k}=\kappa_k$ and $\GG_{k,k}=\wt g_k$, ${\it 0}$ indicates the null matrix, and $\WW$ and $\TT$ are defined in Eq.~\rp{WT}.
Moreover $\vf$ is the vector of input noise operators
$\vf=\ppt{\sqrt{\kappa_1}\,a_1^{in},\dots,\sqrt{\kappa_N}\,a_N^{in},f_1,\dots,f_N,\sqrt{\kappa_1}\,a_1^{in}{\da},\dots, f_1\da,\dots}^T$, with $f_k$ given in Eq.~\rp{fk-}. The correlations of these noise operators can be expressed as
\begin{eqnarray}
\av{\vf_j(t)\ \vf_{j'}(t')}=\delta(t-t')\ \NN_{j,j'}\ ,
\end{eqnarray}
where $\NN$ is the correlation matrix
\begin{eqnarray}
\NN=\ppt{\mmat{cccc}{
{\it 0} & {\it 0}  & \KK  & {\it 0}
\\
{\it 0} & \Xi\al{--} & {\it 0} & \Xi\al{-+}
\\
{\it 0} & {\it 0}  & {\it 0} & {\it 0}
\\
{\it 0} & \Xi\al{+-} & {\it 0} & \Xi\al{++}
}}
\end{eqnarray}
with the matrices $\Xi_{\mu\nu}$ defined in Eq.~\rp{FF}.
The corresponding equation for the $4N\times 4N$ correlation matrix, $\CC$, with elements $\CC_{j,k}=\av{\va_j\ \va_{k}}$, takes the form
\begin{eqnarray}
\dot\CC=\MM\ \CC+\CC\ \MM^T+\NN\ .
\end{eqnarray}
Introducing the linear operator $\LL\CC\equiv\MM\ \CC+\CC\ \MM^T$, we can write the steady state solution as
\begin{eqnarray}\label{Cstfull}
\CC_{st}=-\LL^{-1}\ \NN\ .
\end{eqnarray}
We can also write the $2N\times 2N$ matrix for the steady state correlations, solely for the mechanical degrees of freedom, as
\begin{eqnarray}\label{Cst}
\CC\al{c}_{st}=\ZZ\ \CC_{st}\ \ZZ^T
\end{eqnarray}
where $\ZZ$ is the $2N\times 4N$ matrix
\begin{eqnarray}
\ZZ=\ppt{\mmat{cccc}{
{\it 0} & \id     & {\it 0} & {\it 0}
\\
{\it 0} & {\it 0} & {\it 0} & \id
}}\ .
\end{eqnarray}
The steady state covariance matrix $\VV_{st}\al{c}$, the elements of which are expressed in terms of the vector of operators 
$\vx\al{c}=\ppt{x_1\al{c},\dots x_N\al{c},p_1\al{c},\dots p_N\al{c}}^T$
as $\pg{\VV_{st}\al{c}}_{\ell,\ell'}=\ppt{\av{\vx_\ell\al{c}\ \vx_{\ell'}\al{c}}_{st}+\av{\vx_{\ell'}\al{c}\ \vx_\ell\al{c}}_{st}}/2$, can be written as
\begin{eqnarray}\label{Vst}
\VV_{st}\al{c}=\RR\ \frac{\CC_{st}\al{c}+\CC_{st}\al{c}{}^T}{2}\ \RR^T \ ,
\end{eqnarray} 
where
\begin{eqnarray}
\RR=\ppt{\mmat{cc}{
\id & \id \\
-i\ \id & i\ \id
}}
\end{eqnarray} 
such that $\vx\al{c}=\RR\,\va$.

Finally the corresponding covariance matrix for the original modes $\VV_{st}\al{b}$ can be found
introducing the Bogoliubov matrix $\BBB$ which transforms the field operators $c_j$ and $c_j\da$  into $b_j$ and $b_j\da$, i.e. $b_j=\sum_{j'}\ppt{\BBB_{j,j'}c_{j'}+\BBB_{j,j'+N}c_{j'}\da}$ and $b_j\da=\sum_{j'}\ppt{\BBB_{j+N,j'}c_{j'}+\BBB_{j+N,j'+N}c_{j'}\da}$, that is [see Eqs.~\rp{UbUc} and \rp{UdabU-}]
\begin{eqnarray}
\BBB=\ppt{\mmat{cc}{
\XX\da  & -\YY^T
\\
-\YY\da & \XX^T 
}}\ .
\end{eqnarray}
We find
\begin{eqnarray}
\VV_{st}\al{b}=\RR\ \BBB\ \frac{\CC_{st}\al{c}+\CC_{st}\al{c}{}^T}{2}\ \BBB^T\ \RR^T \ .
\end{eqnarray}

%%%%%%%%%%%%%%%%%%%%%%%%%%%%%%%%%%%%%%%%%%%%%%%%%%%%%%%%%%%%%%			

\end{document}